\input{aipcheck}

\documentclass[
    ,final            
  ]
  {aipproc}

\layoutstyle{6x9}

\begin{document}

\title{High-energy emission from off-axis relativistic jets}

\author{E.V. Derishev}{
  address={Institute of Applied Physics, 46 Ulyanov st., 603950 Nizhny
Novgorod, Russia} }

\author{F.A. Aharonian}{
  address={Max-Planck-Institut f\"ur Kernphysik,
Saupfercheckweg 1, D-69117 Heidelberg, Germany} }

\author{Vl.V. Kocharovsky}{
  address={Institute of Applied Physics, 46 Ulyanov st., 603950 Nizhny
Novgorod, Russia} }

\begin{abstract}

We analyze how the spectrum of synchrotron and inverse Compton
radiation from a narrow relativistic jet changes with the
observation angle. It is shown that diversity of acceleration
mechanisms (in particular, taking the converter mechanism
(Derishev et~al. 2003) into account) allows for numerous
modifications of the observed spectrum. In general, the off-axis
emission in GeV-TeV energy range appears to be brighter, has a
much harder spectrum and a much higher cut-off frequency compared
to the values derived from Doppler boosting considerations alone.
The magnitude of these effects depends on the details of particle
acceleration mechanisms, what can be used to discriminate between
different models.

One of the implications is the possibility to explain
high-latitude unidentified EGRET sources as off-axis but otherwise
typical relativistic-jet sources, such as blazars. We also discuss
the broadening of beam pattern in application to bright transient
jet sources, such as Gamma-Ray Bursts.

\end{abstract}

\maketitle

\section{Introduction}

Relativistic flows are attractive as a solution of the compactness
problem for bright and rapidly variable sources, such as Active
Galactic Nuclei (AGNs) and Gamma-Ray Bursts (GRBs). Lorentz
boosting lowers both the required comoving number density and the
required energy of individual photons, so that the pair-production
opacity can be made smaller than one for a source of given size.
Often (if not always), the relativistic outflows are in the form
of narrow jets, as seen in many AGNs (e.g., Urry \& Padovani 1995)
and deduced theoretically for GRBs (e.g., M\'{e}sz\'{a}ros 2002).

The central dogma for narrow and highly relativistic jets is that
they are bright when viewed head-on, but hardly observable
off-axis (at the viewing angle much larger than $1/\Gamma$, where
$\Gamma$ is the Lorentz factor of a jet) since they appear far
dimmer and strongly redshifted as compared to the head-on
emission. Indeed, due to aberration of light, isotropic in the jet
frame radiation is beamed in the laboratory frame, so that it is
essentially confined within an angle $\sim 1/\Gamma$ from the
jet's velocity. This follows from the equation relating polar
angles $\theta$ and $\theta^{\prime}$ of a light beam in the two
frames (hereafter, prime denotes the comoving-frame quantities):
\begin{equation}
\label{aber} \cos \theta = \frac{\beta - \cos \theta^{\prime}} {1
- \beta \cos \theta^{\prime}},
\end{equation}
where $\beta$ is the jet velocity divided by the speed of light
$c$, the polar axes are directed along the jet's velocity in the
laboratory frame and counter to it in the jet comoving frame: the
latter choice makes Eq.~(\ref{aber}) symmetric with respect to
$\theta$ and $\theta^{\prime}$. The frequencies of emission are
related as
\begin{equation}
\label{freq} \nu = \Gamma \left( 1 - \beta \cos \theta^{\prime}
\right) \nu^{\prime} \equiv  \delta \nu^{\prime}\, .
\end{equation}
Here $\delta = \Gamma (1 - \beta \cos \theta^{\prime})$ is the
Doppler factor. The observed spectral intensity, i.e., the energy
flux per unit solid angle and per unit frequency, is given by
\begin{equation}
\label{bri} I (\nu, \theta) = \delta^n I^{\prime} (\nu^{\prime},
\theta^{\prime})\, ,
\end{equation}
where $n=2$ for a continuous jet and $n=3$ for a relativistically
moving blob (Lind \& Blandford 1985). Below we consider only the
continuous jet case. It is convenient to write the asymptotic form
of Eqs.~(\ref{aber}), (\ref{freq}) and (\ref{bri}) for the case,
where $\theta, \theta^{\prime} \ll 1$:
\begin{equation}
\begin{array}{l}
\displaystyle \nu \simeq \frac{\Gamma \theta^{\prime\, 2}}{2}\,
\nu^{\prime}\, ,\qquad \ \displaystyle \theta \theta^{\prime}
\simeq \sqrt{\frac{8(1-\beta)}{\beta}} \simeq \frac{2}{\Gamma},\\
\displaystyle I \left( \nu, \theta \right) \simeq
\frac{4}{\Gamma^2 \theta^4}\, \, I^{\prime} \left( \frac{\Gamma
\theta^2}{2}\, \nu, \frac{2}{\Gamma \theta} \right).
\end{array}
\end{equation}

For isotropic in the jet frame emission, i.e., for $I^{\prime}
(\nu^{\prime}, \theta^{\prime}) = I^{\prime} (\nu^{\prime})$, the
bolometric brightness drops very rapidly with increase of the
viewing angle: $\displaystyle \int I\, {\rm d}\nu \propto
\theta^{-6}$ as long as $\theta \gg 1/\Gamma$. The situation is
somewhat complicated if the jet is observed at different angles in
the same (narrow) spectral range: hard spectra partially
compensate for the decrease of brightness due to frequency shift,
but the spectra must be unreasonably hard to reverse the general
trend. There is the only way for an off-axis jet not to obey the
dogma about darkening and reddening with increasing observation
angle: this is possible if the emission produced in the jet has
strong anisotropy in the comoving frame.

\section{Anisotopy in the jet frame: conditions and parameters}

The photons produced by a relativistic particle continue to stream in
the direction of the particle's momentum, hence -- in order to have
anisotropic emission -- the same degree of anisotropy is required for
the particle distribution. The origin of every photon can eventually
be traced to a charged particle, so that the anisotropic distribution
of radiating particles means that there is a current in the jet's
plasma, which should quickly decay due to numerous instabilities.
Two conditions have to be met to maintain the anisotropy: there must
be a continuous anisotropic supply of relativistic particles and
these particles must have a lifetime (with respect to radiative
losses) shorter than the isotropisation timescale.

The first condition is almost automatically satisfied at the shock
front for the particles advected with the upstream fluid, whereas
the second requirement precludes production of such particles via
diffusive shock acceleration. Indeed, the diffusive acceleration
proceeds through multiple scattering of the accelerated particle
downstream and the resulting passages of the shock in direction
from downstream to upstream. In the case of relativistic shock,
the particle increases its energy by a factor $\sim 2$ in each
shock passage (e.g., Achterberg et~al. 2001). The particle should
be able to preserve at least a half of its energy over one mean
free path to keep accelerating. So, the distribution of
diffusively accelerated particles is always close to isotropic in
the jet frame.

Anisotropic distribution of radiating particles can be realized in
two ways. One is to have these particles pre-injected upstream, as
in the case of $e^{-}e^{+}$-pair loading ahead of GRB shocks
(Madau \& Thompson 2000). Another, more general way, is to produce
them via non-diffusive converter acceleration mechanism (Derishev
et~al. 2003), which essentially requires isotropisation of
accelerated particles upstream each time before they cross the
shock. In what follows, we take an isotropic distribution of
particles upstream, created in either way. This distribution is
assumed to be a power-law of the particle's energy $\varepsilon$
(${\rm d}N/{\rm d} \varepsilon \propto \varepsilon^{-s}$) with the
cut-off at $\varepsilon_m$. In the comoving frame, this transforms
into the power-law injection at the shock with the same index and
the cut-off at $\varepsilon_m^{\prime} \simeq \Gamma
\varepsilon_m$. Generalization for the case of arbitrary
distribution is straightforward; we skip it for the sake of
brevity.

Let $d(\varepsilon^{\prime})$ be the particle's mean free path and
$\ell (\varepsilon^{\prime})$ its radiation length, both measured
in the jet's frame. In the majority of cases, the function $d
(\varepsilon^{\prime})$ allows power-law representation:
\begin{equation}
d (\varepsilon^{\prime}) \propto \left(
\varepsilon^{\prime}/\varepsilon_{\rm cr}^{\prime} \right)^{p} ,
\end{equation}
where the critical energy $\varepsilon_{\rm cr}^{\prime}$ is
defined from equality $d(\varepsilon_{\rm cr}^{\prime}) = \ell
(\varepsilon_{\rm cr}^{\prime})$. For instance, $p=1$ for a
quasi-uniform magnetic field, that includes the case of Bohm
diffusion, and $p=2$ for the case of small-angle scattering. We
also introduce the characteristic width of the particle
distribution downstream,
\begin{equation}
\label{wid} \phi^{\prime} (\varepsilon^{\prime}) = \left\{
\begin{array}{ll}
1\, , &  \varepsilon^{\prime} < \varepsilon_{\rm cr}^{\prime}\\
\displaystyle \left( \frac{\ell}{d} \right)^{1/p} \propto
\frac{\ell^{1/p}}{\varepsilon^{\prime}}, \qquad & \varepsilon_{\rm
cr}^{\prime} < \varepsilon^{\prime} < \varepsilon_{\rm cr\,
2}^{\prime}\\ \displaystyle 1/\Gamma\, , & \varepsilon^{\prime} >
\varepsilon_{\rm cr\, 2}^{\prime}
\end{array}
\right. \qquad ,
\end{equation}
where the second critical energy $\varepsilon_{\rm cr\,
2}^{\prime}$ is defined by equality $\phi^{\prime}
(\varepsilon_{\rm cr\, 2}^{\prime}) = \Gamma^{-1}$. The equation
(\ref{wid}) has simple physical meaning. The sub-critical
particles ($\varepsilon^{\prime} < \varepsilon_{\rm cr}^{\prime}$)
have enough time to get fully isotropized, whereas for
super-critical ones the width of the distribution function is
equal to their maximum deflection angle $\phi^{\prime}$. Above
$\varepsilon_{\rm cr\, 2}^{\prime}$ the particles lose energy
before being significantly deflected, and the distribution
function preserves its intrinsic width $1/\Gamma$, which
originates due to the Lorentz transformation from the upstream
frame.

Consider some representative situations.

{\bf Synchrotron or self-Compton radiation in the Thomson regime.}
The typical frequency of emitted photons is $\nu^{\prime} \propto
\varepsilon^{\prime\, 2}$. If these types of emission prevail in
radiative losses, then $\ell \propto 1/\varepsilon^{\prime}$.
Consequently, $\phi^{\prime} \propto \varepsilon^{\prime\,
-\frac{p+1}{p}}$, that is  $\phi^{\prime} = (\varepsilon_{\rm
cr}^{\prime}/\varepsilon^{\prime})^{2}$ for the Bohm diffusion.

{\bf Self-Compton radiation in the Klein-Nishina regime.} The
typical frequency of emitted photons is $\nu^{\prime} \propto
\varepsilon^{\prime}$. In the case where such losses are
prevalent, and the spectrum of radiation being comptonised is a
power-law $F_{\nu}^{\prime} \propto \nu^{\prime\, q}$, where $-1 <
q<1$, we have $\ell \propto \varepsilon^{\prime\, q}$. The width
of the particle distribution is $\phi^{\prime} \propto
\varepsilon^{\prime\, \frac{q-p}{p}}$; for the Bohm diffusion
$\phi^{\prime} = (\varepsilon^{\prime}/\varepsilon^{\prime}_{\rm
cr})^{q-1}$.

{\bf Comptonization of external radiation with a logarithmically
narrow spectrum in the Thomson regime.} The frequency of emitted
photons is given by $\nu^{\prime} \propto \varepsilon^{\prime\, 2}
\phi^{\prime\, 2}$. If this is the prevalent type of losses, then
$\ell \propto \left( \varepsilon^{\prime} \phi^{\prime\, 4}
\right)^{-1}$. Thus, $\phi^{\prime} \propto \varepsilon^{\prime\,
-\frac{p+1}{p+4}}$, i.e., $\phi^{\prime} =
(\varepsilon^{\prime}/\varepsilon_{\rm cr}^{\prime})^{-2/5}$ in
the case of Bohm diffusion.

To give an idea of energy scale, the critical energy for
electrons, whose acceleration is limited by the synchrotron
losses, is
\begin{equation}
\varepsilon_{\rm cr}^{\prime} = \frac{3}{2} \frac{ \left( m_e c^2
\right)^2}{\sqrt{e^3 B^{\prime}}}\, ,
\end{equation}
and the associated cut-off frequency of their synchrotron emission
is
\begin{equation}
\label{maxsy} \nu_{\rm cr}^{\prime}  \simeq \frac{0.5}{\pi}
\frac{e B^{\prime}}{m_e c} \left( \frac{\varepsilon_{\rm
cr}^{\prime}}{m_e c^2} \right)^2, \qquad h \nu_{\rm cr}^{\prime}
\simeq 270\, m_e c^2\, .
\end{equation}
The second critical energy is $\Gamma^{\frac{p}{p+1}}$ times
larger than $\varepsilon_{\rm cr}^{\prime}$.

\section{Off-axis spectra and luminosity}

The majority of particles injected at the shock penetrate deep
into downstream, loose energy through synchrotron and other types
of radiation, and form a cooling distribution ${\rm d} N^{\prime}/
{\rm d} \varepsilon^{\prime} = \dot{N^{\prime}}/
\dot{\varepsilon^{\prime}} \propto \ell \varepsilon^{-s}$. The
angle-averaged brightness in the comoving frame can be derived in
a straitforward way:
\begin{equation}
\bar{I^{\prime}} (\nu^{\prime}) \propto
\dot{\varepsilon^{\prime}}\, \, \frac{{\rm d} N^{\prime}}{{\rm d}
\varepsilon^{\prime}} \left( \frac{{\rm d} \nu^{\prime}}{{\rm d}
\varepsilon^{\prime}} \right)^{-1} \propto \nu^{\prime\,
\frac{2-s-x}{x}}\, ,
\end{equation}
where we assumed that the spectrum of an individual particle is
monochromatic with frequency $\nu^{\prime}  \propto
\varepsilon^{\prime\, x}$. The index $x$ is different for
different emission types, as discussed in the end of the previous
section.

Taking into account that emission produced by the super-critical
particles is concentrated within a cone with opening angle
$\phi^{\prime} (\varepsilon^{\prime})$, we find the observed
brightness at small angles $1/\Gamma \ll \theta \ll 1$:
\begin{equation}
I (\nu, \theta) \propto \frac{\delta^2}{\phi^{\prime\, 2}} \left(
\frac{\nu}{\delta} \right)^{\frac{2-s-x}{x}} \simeq
\frac{1}{\theta^2} \left( \frac{\nu}{\delta}
\right)^{\frac{2-s-x}{x}} \, ,
\end{equation}
where $\theta$ and $\phi^{\prime}$ are related as $\theta \simeq
2/(\Gamma \phi^{\prime})$ and $\nu /\delta$ is the function of
$\phi^{\prime}$.

All the changes in the spectrum as compared to the head-on
emission are due to the factor $\phi^{\prime\, -2}$, which is a
rising function of frequency in the range between $\nu_{\rm cr} =
\delta \nu_{\rm cr}^{\prime}$ and the cut-of $\nu_{\rm m} = \left[
\varepsilon^{\prime}(\phi^{\prime})/\varepsilon_{\rm
cr}^{\prime}\right]^x \nu_{\rm cr}$. The difference between
spectral indices in the head-on and off-axis emission depends on
the type of this emission, but the off-axis spectrum is always
much harder.

As an example, let us consider the synchrotron emission in some
detail. In this case, the spectral index between $\nu_{\rm cr}$
and $\nu_{\rm m}$ increases by 2 and 1.5 for the Bohm diffusion
and the small-angle scattering, respectively. For an injected
particle distribution with indices $s$ smaller than $13/3$ or
$10/3$ (the Bohm diffusion and the small-angle scattering,
respectively), that includes the typical assumption $s \simeq 2$,
the resulting spectrum formally appears to be harder than the
low-frequency asymptotic for the synchrotron emission of an
individual particle. This means that in the above frequency range,
as well as immediately below $\nu_{\rm cr}$, the spectrum is
determined by the low-frequency emission of the most energetic
particles.

It is noteworthy that the cut-off frequency of the synchrotron
emission observed at an angle $1/\Gamma \ll \theta \ll 1$ to the
jet's axis can be expressed via the radiative length:
\begin{equation}
\nu_{\rm m} (\theta) \simeq  \nu_{\rm cr} \left( \frac{\ell
(\varepsilon^{\prime})}{\ell (\varepsilon_{\rm cr}^{\prime})}
\right)^{2/p}.
\end{equation}
Here $\varepsilon (\theta)$ is the energy of particles whose
beam-pattern width is equal to $\theta$. In the case of
synchrotron losses, where $\ell \propto \varepsilon^{-1}$, the
dependencies of the observed cut-off frequency on the viewing
angle are $\nu_{\rm m} \propto \phi^{-1}$ for the Bohm diffusion
and $\nu_{\rm m} \propto \phi^{-2/3}$ for the small-angle
scattering. They are much weaker than the relation $\nu_{\rm max}
\propto \phi^{-2}$, which stems from the Doppler factor alone.
Moreover, prevalence of the inverse Compton losses cancels out or
even reverses the above dependencies.

\section{Conclusions and implications}

The jet sources, which are observed off-axis owing to the effect
of beam pattern broadening should exhibit very hard spectra.
Indeed, the emission with broadened beam pattern is produced by
high-energy particles which cool radiatively over a distance
smaller than their mean free path. For such particles the
deflection angle (and hence the width of the beam pattern) is a
function of their energy. An observer situated at a large angle to
the jet axis effectively sees the particle distribution devoid of
its low-energy part, whose emission can only be seen at smaller
viewing angles. Therefore, the off-axis emission is the hardest
possible -- it is essentially as hard as the spectrum of an
individual particle.

The effect of beam-pattern broadening opens an interesting
possibility for observation of the off-axis blazars and GRBs. The
harder spectrum at large viewing angles can explain the phenomenon
of unidentified gamma-ray sources and may even give rise to orphan
GRB afterglows in the TeV range. In addition, a
population-synthesis survey of off-axis sources will be able to
reveal the details and relative importance of the electron cooling
processes.

\begin{theacknowledgments}
E.V. Derishev acknowledges the support from the Russian Science
Support Foundation. This work was also supported by the RFBR grant
no. 02-02-16236, the President of the Russian Federation Program
for Support of Leading Scientific Schools (grant no.
NSh-1744.2003.2), and the program "Nonstationary Phenomena in
Astronomy" of the Presidium of the Russian Academy of Science.
\end{theacknowledgments}

\end{document}